\documentclass[aps,twocolumn]{revtex4}
\usepackage{amsfonts}
\usepackage{amsmath}
\usepackage{amssymb}
\usepackage{graphicx}

\begin{document}
\preprint{ }
\title{Determination of the resistivity anisotropy of SrRuO$_{3}$ by measuring the planar Hall effect}
\author{Isaschar Genish}
\author{Lior Klein}
\affiliation{Department of Physics, Bar-Ilan University, Ramat-Gan
52900, Israel}
\author{James W. Reiner}
\altaffiliation{Present address: Department of Applied Physics,
Yale University, New Haven, Connecticut 06520-8284}
\author{M. R. Beasley}
\affiliation{T. H. Geballe Laboratory for Advanced Materials, Stanford University,
Stanford, California 94305}
\keywords{}%

\begin{abstract}
We have measured the planar Hall effect in epitaxial thin films of
the itinerant ferromagnet  SrRuO$_{3}$ patterned with their
current paths at different angles relative to the crystallographic
axes. Based on the results,  we have determined that SrRuO$_{3}$
exhibits small resistivity anisotropy in the entire temperature
range of our measurements  (between 2 to 300 K); namely, both
above and below its Curie temperature (~150 K). It means that in
addition to anisotropy related to magnetism, the resistivity
anisotropy of SrRuO$_{3}$ has an intrinsic, nonmagnetic source. We
have found that the two sources of anisotropy have competing
effects.

\end{abstract}




\maketitle

\section{Introduction}

The itinerant ferromagnet SrRuO$_{3}$ has attracted considerable
experimental and theoretical effort for its intriguing properties;
including, "bad metal" behavior\cite{trans,nonfermidisorder},
deviation from normal metal optical conductivity \cite{optical}
and negative deviation from Matthiessen's rule
\cite{Matthiessen's}. Consequently, there is growing interest in
obtaining comprehensive characterization of its properties and in
particular its transport behavior. Here we address a basic
transport feature of SrRuO$_{3}$ as we use planar Hall effect
measurements to determine quantitatively the existence of
spontaneous resistivity anisotropy.

Being almost cubic, the few indications for spontaneous
resistivity anisotropy in SrRuO$_{3}$ have been subtle and
qualitative; for example, resistivities measured along [001] and
[1$\bar{1}$0] exhibit different critical behavior and they also
seem to have slightly different values in the paramagnetic state
\cite{trans}. These small differences imply that even if the
anisotropy is real (and not due to spurious effects) it does not
exceed few percents; hence, its quantitative determination is
challenging.

In principle, obtaining quantitative characterization of
spontaneous resistivity in a conductor can be achieved by direct
measurements of the resistivity for different current directions.
This method, however, is not very useful when the anisotropy is on
the order of few percents, since it is affected by other sources
(e.g., variations in geometrical factors and in the number and
type of defects) whose contribution could be on the order of that
of the intrinsic anisotropy.
 To overcome this
difficulty, we deduce the intrinsic spontaneous resistivity
anisotropy in SrRuO$_{3}$ by measuring the planar Hall effect
(PHE) of this compound.

The PHE \cite{PHE1,PHE2,amr} is the appearance of transverse
resistivity, $\rho_{xy}$, observed in conductors with resistivity
anisotropy. In magnetic compounds it is usually related to the
anisotropic magnetoresistance \cite{amr} which is the dependence
of the longitudinal resistivity, $\rho_{xx}$, on the angle
$\theta$ between the current and the internal magnetization. For
many compounds it is found that:

$\rho_{xx}=\rho_\bot+(\rho_\|-\rho_\bot)\cos^2\theta$ and

$\rho_{xy}=(\rho_\|-\rho_\bot)\cos\theta\sin\theta$

where $\rho_\|$ and $\rho_\bot$ are the resistivities with the
magnetization parallel and perpendicular to the current,
respectively. However, simply measuring the transverse resistivity
is also not sufficient since misalignment of contact leads would
introduce a regular longitudinal contributions that is hard to
distinguish from the transverse signal, since contrary to ordinary
and extraordinary Hall effects, the planar Hall effect is
symmetric under magnetization reversal, as is the longitudinal
resistivity. Therefore, for reliable determination of the PHE one
needs to demonstrate the expected dependence of the PHE on the
angle between the current path and the principal axes of the
resistivity tensor.

Despite being commonly associated with magnetism PHE is a
phenomenon which  arises whenever there is resistivity anisotropy
and the current is not along one of the principal axes of the
resistivity tensor; hence, it can be used to directly determine
the intrinsic anisotropy of a conductor irrespective of the source
of the anisotropy, including a non-magnetic source. However, if
non-magnetic source of the anisotropy exists, one cannot rotate
the direction of the principal axes of the resistivity tensor by
changing the direction of an applied magnetic field. Therefor, we
"rotate" the current path relative to the crystallographic
 axes by patterning current paths in different directions.
 Using this method, we have quantitatively determined the
 resistivity anisotropy of SrRuO$_{3}$ both
 above and below the Curie temperature and we have found that there are both magnetic and non-magnetic sources
 to the anisotropy with competing effects.  The success of this method indicates that it could be
 a useful method for determining subtle anisotropy of other conductors, as well.

\section{Measurements and discussion}

Our samples are epitaxial films of SrRuO$_{3}$ grown on slightly
miscut (2$^{\circ}$) SrTiO$_{3}$. The films are orthorhombic
($a=5.53$ \AA , $b=5.57$ \AA , $c=7.85$ \AA) \cite{Marshall} and
their Curie temperature is $\sim150$ K. The films grow with the
in-plane $c$ axis perpendicular to the miscut, and $a$ and $b$
axes at 45 degrees relative to the film plane. The films exhibit
uniaxial magnetocrystalline anisotropy with the easy axis close to
the $b$ axis \cite{Marshall}. These are twin-free films whose high
quality has been previously manifested, including in exhibiting
low-temperature magnetoresisitance quantum oscillations
\cite{oscillations}.

The data presented here are from a 27 nm thick film with
resisitivity ratio of $\sim12$ on which there are 8 identical
patterns, each along a different crystallographic direction in the
(110) plane. The angles of the patterns relative to the
[1$\bar{1}$0] axis (in the film plane) are: -45 ,0, 15, 30, 45,
60, 75 and 90 degrees. Each pattern allows longitudinal and
transverse resistivity measurements (see Figure 1).

\begin{figure}[ptb]
\includegraphics[scale=0.4, trim=150 350 200 -75]{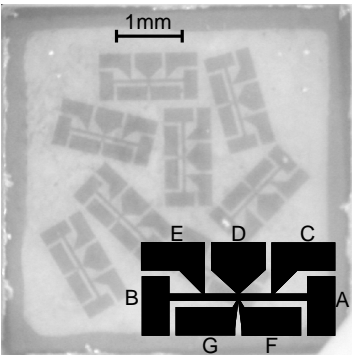}
\caption{Photo of the measured sample which consists of 8 patterns
at different orientations. Inset: A scheme of an individual
pattern.}
 \label{fig1}
\end{figure}

The resistance measured between the pads marked F and D, $R_{FD}$,
has several contributions: a) longitudinal resistance due to
longitudinal shift between the two leads  b) ordinary Hall effect
[OHE] and extraordinary Hall effect [EHE] \cite{EHE_s2} due to
magnetic field and magnetization, and c) PHE. According to the
pattern design, the longitudinal resistance should be half the
resistance between F and G, $R_{FG}$; therefore, it can be
subtracted. The OHE and EHE are antisymmetric signals; hence, they
are determined by interchanging current and voltage leads
\cite{rec}.
Figure 2a. shows  R$_{FD}$ as a function of temperature for all
the patterns. The measurements are at zero applied field after
field cooling to avoid contribution of magnetic domain walls to
resisitivity \cite{DWR_m}.  The antisymmetric contribution to
R$_{FD}$ (obtained by current and voltage interchange) for all the
patterns is shown in Figure 2b. Since the applied field is zero,
the obtained signal is the EHE. We see that this contribution is
almost identical for all patterns (and consistent with previous
reports \cite{eheM}) indicating that,  as expected, only the
perpendicular component of the magnetization, which is identical
for all patterns, is relevant. In Figure 2c we show the PHE for
all the patterns after subtracting the antisymmetric part and the
expected contribution due to longitudinal resistance.

\begin{figure}[ptb]
\includegraphics[scale=0.5, trim=250 0 200 150]{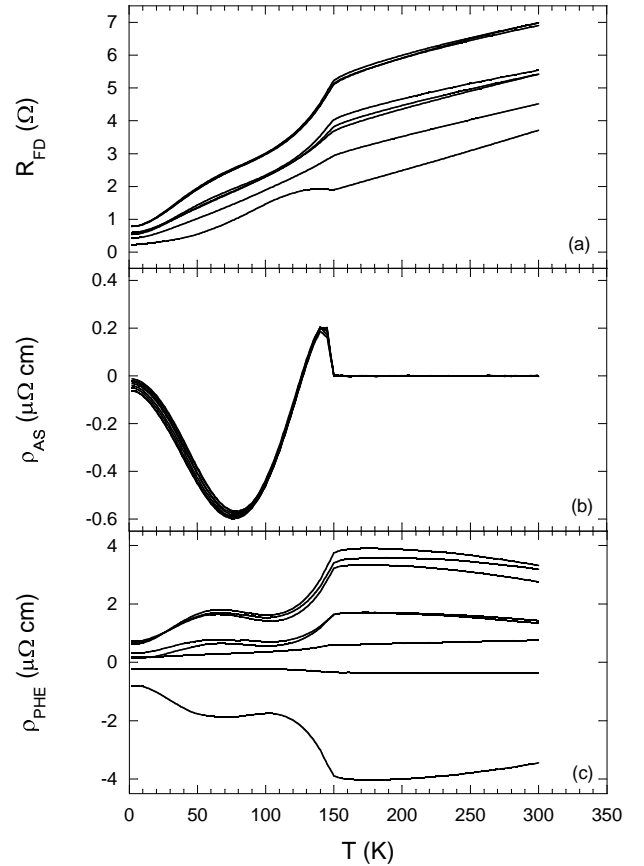}
\caption{a)  $R_{FD}$ as a function of temperature for the 8
patterns. In decreasing order of $R_{FD}$ value at 300 K the
curves correspond to patterns at angles of : 45, 30, 15, 0, 75 90
and -45 degrees b) The antisymmetric contribution to $R_{FD}$. c)
PHE contribution to $R_{FD}$. In decreasing order of their value
at 300 K the curves correspond to patterns at angles of : 45, 60,
30, 15, 75,  0, 90 and -45 degrees.}
 \label{fig2}
\end{figure}

We note that the PHE curves exhibit similar qualitative angular
dependence with the largest PHE  obtained for $\theta = \pm
45^\circ$ and the smallest  for $\theta = 0^\circ, 90^\circ$. This
indicates that the resistivities in the $0^\circ$ and $90^\circ$
directions, $\rho_0$ and $\rho_{90}$,  are the
 principal axes of resistivity and we expect
that at each temperature the PHE for current in the $\theta$
direction will obey
$\rho_{xy}=(\rho_{0}-\rho_{90})\sin\theta\cos\theta$. It also
means that between any two PHE curves obtained for patterns at
angles $\theta_1$ and $\theta_2$ there will be
temperature-independent proportionality given by
$\cos\theta_1\sin\theta_1/ \cos\theta_2\sin\theta_2$.

As noted above, the longitudinal resistance contribution to
R$_{FD}$ should be $a\times R_{FG}$ with $a=0.5$. However, minute
lithography variations may slightly change this factor.
 If the correct longitudinal contribution is precisely
determined for all angles, we expect that all the PHE curves will
be proportional to each other. Therefore, we use MATLAB for
refining "$a$" within few percents in order to obtain the
simultaneously best proportionality between the data sets of PHE
obtained for the different angles. The process does not impose the
magnitude of the proportionality factor. Figure 3a. shows the PHE
for all the angles after the refinement. To show the
proportionality we present in Figure 3b the PHE at various angles
as a function of the PHE obtained for $\theta=45^\circ$. The inset
of Figure 3b shows the proportionality factor as a function of the
pattern angle where the solid line is sin$\theta$cos$\theta$ (with
no adjustable parameters). This remarkable consistency obtained
between all PHE curves strongly supports the reliability of the
obtained PHE curve.

\begin{figure}[ptb]
\includegraphics[scale=0.45, trim=190 -20 200 100]{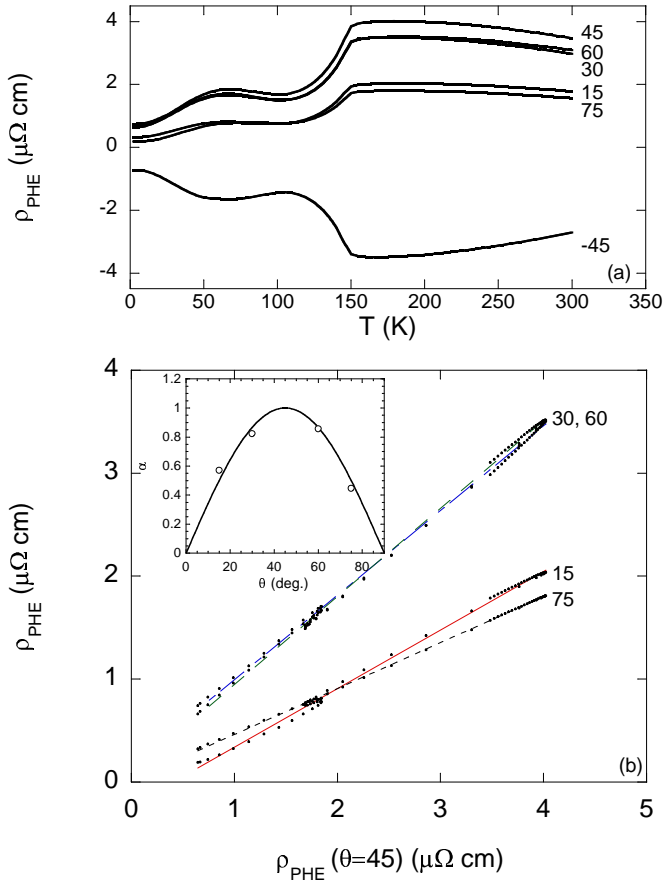}
\caption{a) Refined PHE (see text) vs temperature. The number near
the curve indicates the pattern orientation. b) The PHE in
patterns oriented at various angels vs the PHE in the pattern
oriented at 45 degrees. The lines are linear fits. Inset: The
slope of the linear fits , $\alpha$, as a function of $\theta$.
The line is the expected sin$\theta$cos$\theta$ dependence.}
\label{fig3}
\end{figure}

Figure 4 presents the resistivity anisotropy as obtained from the
PHE. The results show that in the entire temperature range of our
measurements (2 to 300 K) $\rho_0$  (along [110]) is larger than
$\rho_{90}$ (along [001]). As a consistency check we show that if
we allow for small adjustments of few percents, we can correlate
the PHE with the difference between $\rho_{xx}$ measured at
different angles. The agreement is very satisfactory in view of
the effect of other sources on the variations between different
patterns, such as different defect concentration. It should be
noted that only the fact that we have determined the anisotropy by
PHE measurements allowed us to make the small adjustments in the
longitudinal resistivities.

\begin{figure}[ptb]
\includegraphics[scale=0.55, trim=170 250 200 300]{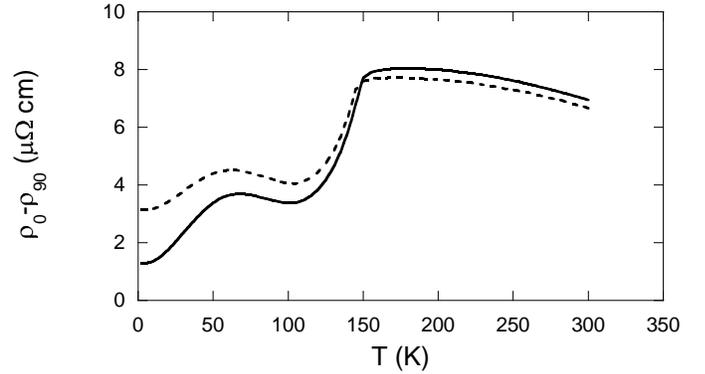}
\caption{The anisotropy $\rho_{0}-\rho_{90}$ as determined by PHE
measurements (continues) and from subtraction of longitudinal
resistivities (dashed).} \label{fig4}
\end{figure}

An interesting and somewhat surprising observation is that the
anisotropy persists in the paramagnetic state. Despite the lack of
long range order in the paramagnetic state one cannot exclude a
priori magnetic origin for the paramagnetic anisotropy in view of
the observed anisotropic paramagnetic susceptibility \cite{large}.
To address possible magnetic origin we examine the dependence of
the anisotropy on magnetization in the paramagnetic state.

\begin{figure}[ptb]
\includegraphics[scale=0.5, trim=100 140 100 250]{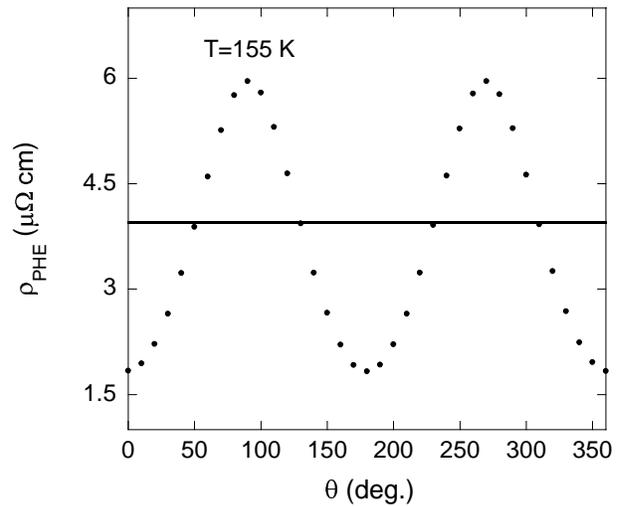}
\caption{The PHE for the pattern oriented at 45 degrees at 155 K
as a function of $\theta$, the angle between an in-plane 8 T field
and [1$\bar{1}$0]. The dashed line is the zero field value of the
PHE.} \label{fig5}
\end{figure}

Figure 5 shows the change in PHE (namely the anisotropy) when a
field of 8 T is rotated in the film plane above the Curie
temperature at 155 K. The PHE of the pattern oriented at 45
degrees varies considerably with the rotation of the field.
However, the sign of the anisotropy remains unchanged despite the
fact that the magnetization completes a full rotation. This
clearly indicates that in addition to magnetism there is also
non-magnetic source of the resistivity anisotropy.

Figure 6 shows the effect of applying a magnetic field in the
paramagnetic state along the b axis which is the easy axis of
magnetization. We see that the PHE decreases as the induced
magnetization increases. The inset shows the relative change in
the PHE for all patterns as a function of $H^2$.  We note that all
patterns show the same field dependence, as expected from the fact
that all PHE values are related by temperature-independent and
field-independent constants. At higher temperatures we see also
that the magnetically-related anisotropy is proportional to $H^2$;
namely, to $M^2$. The fact that the induced magnetization
 decreases the PHE indicates that the magnetic and non-magnetic
sources of anisotropy have opposite effects. In view of this
observation, the sharp decrease in the zero-field PHE when
temperature decreases below $T_{c}$ is understood as a result of
the onset of spontaneous magnetization.

\begin{figure}[ptb]
\includegraphics[scale=0.45, trim=100 220 100 300]{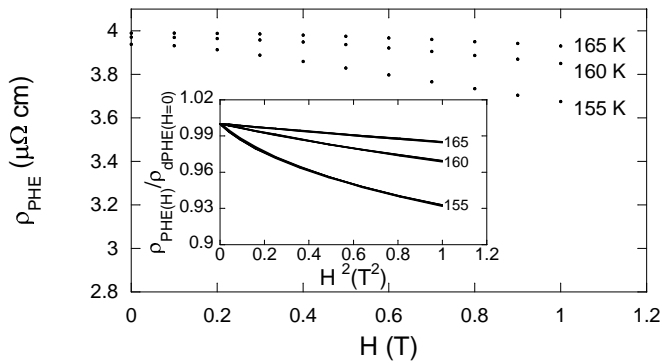}
\caption{a) The PHE for the 45 degrees oriented pattern vs $H$ at
155, 160 and 165 K.  Inset: PHE normalized by its zero-field value
as a function of $H^2$ for all the patterns.} \label{fig6}
\end{figure}

In the paramagnetic state we see a gradual decrease of the PHE
with increasing temperature. The anisotropy is probably related to
the deviation from cubic structure. Studies have shown that
SrRuO$_{3}$ films grown on SrTiO$_{3}$ undergo two structural
phase transitions  orthorhohmbic to tetragonal at
$\sim350^{\circ}$ C and tetragonal to cubic at $\sim600^{\circ}$ C
\cite{transitions}. Therefore, $600^{\circ}$ C is the upper limit
for the observed anisotropy.

\section{conclusions}

By measuring the planar Hall effect on patterns with current paths
oriented in different crystallographic directions we were able to
determine both magnetic and non-magnetic sources of anisotropy
present in epitaxial films of SrRuO$_{3}$. We find the the
in-plane principal axes are [1$\bar{1}$0] and [001]. That
non-magnetic anisotropy makes the resistivity along [1$\bar{1}$0]
larger while spontaneous magnetization (along [010]) decreases
this anisotropy.

\section*{Acknowledgments}

This research was supported by The Israel Science Foundation
founded by the Israel Academy of Sciences and Humanities.


\begin{thebibliography}{9}
\bibitem {trans}L. Klein, J. S. Dodge, C. H. Ahn, G. J. Snyder, T. H. Geballe, M.
R. Beasley, and A. Kapitulnik, Phys. Rev. Lett. \textbf{77}, 2774,
(1996).

\bibitem {nonfermidisorder}M. S. Laad, and E. M\"{u}ller-Hartmann
Phys. Rev. Lett. \textbf{87}, 246402 (2001); Carsten Timm, M. E.
Raikh, and Felix von Oppen, $ibid$. \textbf{94}, 036602, (2005).

\bibitem {optical}P. Kostic, Y. Okada, N. C. Collins, Z. Schlesinger, J. W. Reiner,
 L. Klein, A. Kapitulnik, T. H. Geballe, and M. R. Beasley, Phys. Rev. Lett. \textbf{81}, 2498, (1998);
  J. S. Dodge, C. P. Weber, J. Corson, J. Orenstein, Z. Schlesinger,
J. W. Reiner, and M. R. Beasley, $ibid$. \textbf{85}, 4932–4935
(2000).

\bibitem {Matthiessen's}L. Klein, Y. Kats, N. Wiser, M. Konczykowski, J. W. Reiner, T. H.
Geballe, M. R. Beasley, and A. Kapitulnik
  Europhys. Lett. \textbf{55}, 532 (2001).

\bibitem {PHE1}C. Goldberg and R. E. Davis, Phys. Rev. \textbf{94}, 1121 (1954);
F. G. West, J. Appl. Phys. \textbf{34}, 1171 (1963).

\bibitem {PHE2}W. M. Bullis, Phys. Rev. \textbf{109}, 292 (1958).

\bibitem {amr}T. R. McGuire and R. I. Potter, IEEE Trans. Magn. MAG, \textbf{11}, 1018
(1975).
\bibitem {Marshall}A. F. Marshall, L. Klein, J. S. Dodge, C. H. Ahn, J. W. Reiner,
 L. Mieville, L. Antognazza, A. Kapitulnik, T. H. Geballe, and M. R.
 Beasley,
 J. Appl. Phys. \textbf{85},
4131 (1999).

\bibitem {oscillations}A.P. Mackenzie, J.W. Reiner, A.W. Tyler, L.M. Galvin,
 S.R. Julian, M.R. Beasley, T.H. Geballe, and A. Kapitulnik, Phys. Rev. B \textbf{58}, R13318
 (1998).

\bibitem {large}Y. Kats, I. Genish, L. Klein, J. W. Reiner, and M. R. Beasley, Phys. Rev. B \textbf{71}, 100403 (2005).                                                                                        %

\bibitem {EHE_s2}J. Smit, Physica \textbf{21}, 877 (1955);
J. M. Luttinger, Phys. Rev. \textbf{112}, 739 (1958); J. Smit,
Physica \textbf{24}, 39 (1958). L. Berger, Phys. Rev. B
\textbf{2}, 4559 (1970).

\bibitem {rec}M. B\"{u}ttiker, Phys. Rev. Lett. \textbf{57}, 1761 (1986).

\bibitem {DWR_m}L. Klein, Y. Kats, A. F. Marshall, J. W. Reiner, T. H. Geballe, M. R. Beasley, and A.
Kapitulnik,
  Phys. Rev. Lett. \textbf{84}, 6090 (2000); M. Feigenson,  L. Klein, J. W. Reiner, and M. R.
Beasley, Phys. Rev. B \textbf{67}, 134436 (2003).

\bibitem {eheM}L. Klein, J. R. Reiner, T. H. Geballe, M. R. Beasley, and A. Kapitulnik, Phys. Rev. B \textbf{61},
 R7842 (2000); Z. Fang, N. Nagaosa, K. S. Takahashi, A. Asamitsu, R. Mathieu, T. Ogasawara, H. Yamada, M. Kawasaki, Y. Tokura,
  and K. Terakura, Science \textbf{302}, 92 (2003).

\bibitem {transitions}J. P. Maria, H. L. McKinstry, and S. Trolier-McKinstry, Appl. Phys. Lett. \textbf{76}, 3382
(2000).


\end{thebibliography}
\end{document}